\documentclass[12pt]{article}
\usepackage[cp866]{inputenc}

\usepackage{amsfonts,amsmath,amssymb}
\usepackage{amssymb}

\begin{document}\def\p{\phi}\def\P{\Phi}\def\a{\alpha}\def\e{\epsilon}
\def\be{\begin{equation}}\def\ee{\end{equation}}\def\l{\label}
\def\0{\setcounter{equation}{0}}\def\b{\beta}\def\S{\Sigma}\def\C{\cite}
\def\r{\ref}\def\ba{\begin{eqnarray}}\def\ea{\end{eqnarray}}
\def\n{\nonumber}\def\R{\rho}\def\X{\Xi}\def\x{\xi}\def\la{\lambda}
\def\d{\delta}\def\s{\sigma}\def\f{\frac}\def\D{\Delta}\def\pa{\partial}
\def\Th{\Theta}\def\o{\omega}\def\O{\Omega}\def\th{\theta}\def\ga{\gamma}
\def\Ga{\Gamma}\def\t{\times}\def\h{\hat}\def\rar{\rightarrow}
\def\vp{\varphi}\def\inf{\infty}\def\le{\left}\def\ri{\right}
\def\foot{\footnote}\def\vep{\varepsilon}\def\N{\bar{n}(s)}
\def\k{\kappa}\def\sq{\sqrt{s}}\def\bx{{\mathbf x}}\def\La{\Lambda}
\def\bb{{\bf b}}\def\bq{{\bf q}}\def\cp{{\cal P}}\def\tg{\tilde{g}}
\def\cf{{\cal F}}\def\bN{{\bf N}}\def\Re{{\rm Re}}\def\Im{{\rm Im}}
\def\bk{\mathbf{k}}\def\cl{{\cal L}}\def\cs{{\cal S}}\def\cn{{\cal N}}
\def\cg{{\cal G}}\def\q{\eta}\def\ct{{\cal T}}\def\bbs{\mathbb{S}}
\def\bU{{\mathbf U}}\def\bE{{\mathbf e}}\def\bc{{\mathbf C}}
\def\vs{\varsigma}\def\cg{{\cal G}}\def\ch{{\cal H}}\def\df{\d/\d }
\def\mz{\mathbb{Z}}\def\ms{\mathbb{S}}\def\kb{{\mathbb
K}}\def\cd{\mathcal D}\def\mj{\mathbf{J}}\def\Tr{{\rm Tr}}
\def\bu{{\mathbf u}}\def\by{{\mathrm y}}\def\bp{{\mathbf p}}
\def\k{\kappa} \def\cz{{\mathcal Z}}\def\ma{\mathbf{A}}
\def\me{\mathbf{E}}\def\mp{\mathbf{P}}\def\ra{\mathrm{A}}
\def\ru{\mathrm{u}}\def\rP{\mathrm{P}}\def\rp{\mathrm{p}}\def\z{\zeta}
\def\my{\mathbf Y}

\begin{center}
{\huge\bf On the connection between classical and quantum
descriptions}

{\large\bf I.D.Mandzhavidze\\ \it Physics Institute, Georgian Academy
of Sciences}

\end{center}

\begin{abstract}
The paper develops the idea that the dynamics of both classical and
quantum processes is time reversible. It is shown how this classical
analogy allows one to define  the measure for the path integral in
quantum mechanics.
\end{abstract}

\section{Introduction}\0

The present approach to the problem of the definition of path
integrals is a generalization of the standard stationary phase
method: the contributions are given by the exact solutions of the
equation \be \d\bar S=0, \l{1.1}\ee where the action $\bar S$
includes a random source of quantum perturbations.

In quantum theories one often encounters problems where the boundary
conditions for Eq.(\r{1.1}) are missing. We shall use an additional
selection rule which amounts to the following: those quantum
processes are important which correspond to the largest classical
measure. The derivation of this rule is the main purpose of the
present paper. To this end (see Ref. \C{aa}) we shall find the
connection with the classical description (Sec.2) and then
(Ref.\C{ab}), using a quantum-mechanical example, shall show the role
played by the classical definition of measure in quantum processes.

The technical aspect of this idea is the suggestion to calculate
directly the probability, which has a classical interpretation,
avoi9ding the intermediate step of the calculations the amplitudes.
in present paper we confine ourselves to the simplest problem - the
motion of one particle in a potential $V(x)$. We shall use the
semiclassical approximation since the result are independent of the
magnitude of the quantum corrections. All our results can be derived
by the methods of quantum mechanics, and the example discussed serves
only as an illustration of our approach.

Let the amplitude $A(x_2,T;x_1,0)$ describe the motion of the
particle from the point $x_1$ to the point $x_2$ during the time $T$.
Using the spectral representation \be
A(x_2,T;x_1,0)=\sum_n\psi_n(x_2)\psi_n^*(x_1)e^{iE_nT},\l{1.2}\ee for
the probability we have: \be W(x_2,T;x_1,0)=\sum_{n_1,n_2}\psi_n(x_2)
\psi_n^*(x_1)\psi_n^*(x_2)\psi_n(x_1)e^{i(E_{n_1}-E_{n_2})T}.\l{1.3}\ee
If one takes into account the condition of orthogonality \be \int dx
\psi_n(x)\psi_m^*(x)=\d_{n,m},\l{1.4}\ee then the quantity \be \int
dx_2dx_1 W(x_2,T;x_1,0)=\sum_n=\int\f{dx dp}{2\pi}=\O^2\l{1.5}\ee is
the independent of $T$ and coincides with the number of states, or,
in the semiclassical approximation, with the volume of the phase
space of one particle. In Sec.2 we show the connection between the
classical and quantum descriptions and derive Eq.(\r{1.5}).

In Sec.3 we consider the role of the conservation laws. Thus, the
Fourier transform of $A(x_2,T;x_1,0)$ with respect to $T$ \be
a(x_2,x_1;E)=\sum_n\f{\psi_n(x_2)\psi_n^*(x_1)}{E-E_n-i0}\l{1.6}\ee
leads to probability \be\o(x_2,x_1;E)=\sum_{n_1,n_2}
\f{\psi_n(x_2)\psi_n^*(x_1)}{E-E_{n_1}-i0} \f{\psi_n^*(x_2)
\psi_n(x_1)}{E-E_{n_2}+i0}.\l{1.7}\ee From this one has $$
\bar\o(E)=\int dx_1dx_2\o(x_2,x_1;E)=\sum_{n}\le|\f{1}{E-E_{n}-i0}
\ri|^2=$$\be= \f{1}{0}\sum_n\Im\f{1}{E-E_{n}-i0}=\O\sum_n\d(E-E_{n}),
\l{1.8}\ee  where $\o=1/0=\infty$. We shall show that this infinite
coefficient is a consequence of the translational invariance of the
problem with respect to time. As expected, the conservation laws
reduce the number of the degrees of freedom, which is reflected in
the different number of zero modes in (\r{1.8}) and (\r{1.5}).

The appearance of infinities in (\r{1.8}) and (\r{1.5}) shows that
the contributions discussed are realized on the largest measure. This
question is discussed more fully in Sec.4. Finally, the results of
the paper are summarized up in Sec.5.

\section{The generalized stationary-phase method}\0

Calculating the quantity \be
|A|^2=\rm<in|out><in|out>^*=<in|out><out|in>,\l{2.1}\ee one can show
that the converging and diverging waves interfere in such a way that
some of the contributions to $|A|^2$ cancel each other \C{aa}. In
order to see this it is convenient to write the amplitude in terms of
path integrals: \be A(x_2,T;x_1,0)=\int^{x(T)=x_2}_{x(0)=x_1}
\f{Dx}{C_T} e^{-iS_T(x)},\l{2.2}\ee where the action $S_T$ is gien by
\be S_T(x)=\int^T_0dt\le(\f{1}{2}~\dot x^2-V(x)\ri)\l{2.3}\ee and
$C_T$ is the standard normalization coefficient: \be C_T=\int^{x(T)=
x_2}_{x(0)=x_1}\exp\{-\f{i}{2}\int^T_0dt~\dot x^2\}\l{2.4}\ee Let us
calculate the quantity \be W(x_2,T;x_1,0)=\int^{x(T)=x_2}_{x(0)=x_1}
\f{Dx_+}{C_T}\f{Dx_-}{C_T^*} e^{-iS_T(x_+)+iS_T(x_-)}\l{2.5}\ee For
simplicity below we assume that in (\r{2.2}) the integration is over
trajectories for which the action is real.

In order to to take into account explicitly the interference between
the contributions of the trajectories $x_+(t)$ and $x_-(t)$ we shall
go over from the integration over two independent trajectories $x_+$
and $x_-$ to the pairs $(x,e)$: \be x_|pm(t)=x(t)\pm e(t).\l{2.6}\ee
Upon this substitution of (\r{2.6}) into (\r{2.5}), the argument of
the exponent takes the form \be S_T(x+e)-S_T(x-e)2\int_0^Tdt e(\ddot
x+V'(x))- \tilde S_T(x,e),\l{2.7}\ee where $\tilde S_T(x,e)$ is the
remainder of the expansion in powers of $e(t)$ ($\tilde S_T=O(e^3)$).
Note that in (\r{2.7}) we have discarded the "surface" term \be
\int_0^Tdt\pa_t(e\dot x)=e(T)\dot x(T)-e(0)\dot x(0)=0,\l{2.8}\ee
since the boundary points of the trajectories $x_+(0)=x_-(0)=x_1$ and
$x_+(T)=x_-(T)=x_2$ are not varied: \be e(o)=e(T)=0.\l{2.9}\ee Next,
\be Dx_+Dx_-=J Dx De,\l{2.10}\ee where $J$ is an unimportant Jacobian
of the transformation.

As a result of the replacement (\r{2.6}) we have \be W(x_2,T;x_1,0)=
\int^{x(T)=x_2}_{x(0)=x_1} \f{Dx}{|C_T|^2}\int^{x(T)=0}_{x(0)=0}
De~\exp\le\{2i\int_0^T dt e(\ddot x+V'(x))+\tilde S_T(x,e)\ri\}.
\l{2.11}\ee One can make use of the formulae \be e^{i\tilde
S_T(x,e)}= \hat e(e',j)e^{i\tilde S_T(x,e')}\exp\{-2i\int_0^Te(t)
j(t)dt\},\l{2.12}\ee where we have introduced the operator \be
e(e',j)=\lim_{e=j=0}exp\{-\f{1}{2i}\int_0^T\f{\d}{\d j(t)} \f{\d}{\d
e(t)}\},\l{2.13}\ee after which from (\r{2.10}) we have found that $$
W(x_2,T;x_1,0)=J\hat e(e',j)\int^{x(T)=x_2}_{x(0)=x_1}
\f{Dx}{|C_T|^2} e^{i\tilde S_T(x,e')}\t$$ $$\t\int^{x(T)=0}_{x(0)=0}
De~\exp\le\{2i\int_0^T dt (\ddot x+V'(x)-j)e\ri\}=$$\be= J'\hat
e(e,j)\int^{x(T)=x_2}_{x(0)=x_1} \f{Dx}{|C_T|^2} e^{i\tilde
S_T(x,e)}\prod_{t\neq 0,T}\d(\ddot x+V'(x)-j),\l{2.14}\ee where the
functional $\d$-function \be \prod_{t\neq 0,T}\d(\ddot x+V'(x)-j)=
\f{J}{J'}\int^{x(T)=0}_{x(0)=0} De~\exp\le\{2i\int_0^T dt (\ddot
x+V'(x)-j)e\ri\}\l{2.15}\ee has arisen as a result of total reduction
of the contributions of the trajectories that are unphysical for the
classical equation of motion \be \ddot x(t)+V'(x)=j(t).\l{2.16}\ee
Note that this equation can be obtained by variation of the effective
action $$\bar S(x)=S_T(x)+\int_0^Tdt x(t)j(t),$$ where $j(t)$ is an
external perturbation force. Following the definition of the operator
(r{2.13}), we must turn out attention to the expansion of the
solutions of the Eq.(\r{2.16}) in powers of $j(t)$. Let us note also
that the operator (\r{2.13}) is Gaussian, so that we can assume that
the system under consideration is perturbed by a random force $j(t)$
(in this connection see \C{ab}).

The qualitative side of the derivation of the exact formulae
(\r{2.14}) is as follows. By virtue of the derivation of the
definition of $W$, the difference $S_T(x_+)-S_T(x_-)$ in (\r{2.5})
coincides with the action during the cycle, so that by definition we
are only interested in reversible processes. Upon the substitution
(\r{2.6}) we have identified the "true" trajectory $x(t)$ and the
virtual deviation $e(t)$. Then the quantity $e(\ddot x+V'(x)-j)$
coincides with virtual work. By contrast to classical mechanics one
has to integrate over $e(t)$, as a result of which the measure of the
remaining path integral takes a Dirac $\d$-function form. In other
words, the proposed definition of the measure of the path integral is
similar to  the classical d'Alembert's principle. As is known, on the
basis of this principle the theory can take into account any external
perturbations [in our case, the perturbation introduced by the source
$j(t)$].

In the semiclassical approximation $\hat e(e,j)$ is given by the
$j\to0$ limit, and thus from (\r{2.14}) we find that \be
W(x_2,T;x_1,0)=J'\hat e(e,j)\int^{x(T)=x_2}_{x(0)=x_1}
\f{Dx}{|C_T|^2}\prod_{t\neq 0,T}\d(\ddot x+V'(x)),\l{2.17}\ee Let the
solution of the homogeneous equation \be \ddot x+V'(x)=0\l{2.18}\ee
be $x_c(t)$, with $x(0)=x_1$ and $x(T)=x_2$. Then \be
W(x_2,T;x_1,0)=J'\hat e(e,j)\int^{x(T)=x_2}_{x(0)=x_1}
\f{Dx}{|C_T|^2}\prod_{t\neq 0,T}\d(\ddot x+V''(x_c)x),\l{2.19}\ee The
remaining integral is calculated by the standard methods \C{ac} (here
it is more convenient to represent (\r{2.19}) as a production of two
Gaussian integrals). As a result we find \be W(x_2,T;x_1,0)=
\f{1}{2\pi}\le|\f{\pa^2 S_T(x_c)}{\pa x_c(0)\pa x_c(T)}\ri|_{x_c(0)=
x_1,x_c(T)=x_2}.\l{2.20}\ee Next, let us recall that the full
derivative of the classical action is \be
dS=p_2dx_2-p_1dx_1,\l{2.21}\ee where $p_2$ and $p_1$ are,
respectively, the final and initial momentum. Then, however, \be
\le|\f{\pa^2 S_T}{\pa x_1\pa x_2}\ri|dx_2=dp_1,\l{2.22}\ee as a
result of which we see that \be\int dx_1 dx_2
W(x_2,T;x_1,0)=\int\f{dx_1 dx_2}{2\pi}=\O^2,\l{2.23}\ee which
coincides with (\r{1.5}).

In the derivation of (\r{2.23}) we have simplified the problem
somewhat by considering a unique solution of Eq.(\r{2.16}). A more
complete solution of the problem is given in the next section. Here
it was only important to demonstrate that the contributions to the
functional integrals are determined by the exact solution of the
classical equation (\r{2.16}), which we interpret as a connection
between the classical and quantum descriptions independently of the
magnitude of the quantum corrections.

\section{Taking the conservation laws into account}\0

Let us consider the motion in the phase space. To this end, we
substitute into (\r{2.2}) the equalities \be1=\int\f{Dp}{B_T}\exp
\le\{-\f{i}{2}\int_0^Tdt(p-\dot z)^2\ri\},~~B_T= \int Dp\exp
\le\{-\f{i}{2}\int_0^Tdt p^2\ri\}.\l{3.1}\ee Then $$
a(x_1,x_2;E)=\int_0^\infty dT e^{-ieT}A(x_2,T;x_1,0)=$$
\be=\int_0^\infty dT e^{-ieT}\int^{x(T)=x_2}_{x(0)=x_1}\f{Dx
Dp}{Z_T^2}\exp\{-iS_T(x,p) \},\l{3.2}\ee where the action $S_T$ is
given by \be S_T(x,p)= \int_0^T dt(p\dot x-H(p,x))\l{3.3}\ee and $H$
is the Hamiltonian \be H(p,x)=1/2p^2+V(x).\l{3.4}\ee The
normalization coefficient is \be Z_T=\int Dx
Dp\exp\le\{-\f{i}{2}\int_0^T(p_2-2p\dot x)\ri\}.\l{3.5}\ee Below we
shall study the simplest example, where the potential has a single
minimum at $x=0$.

Proceeding in the some way as in the derivation of (\r{2.14}), we
calculate $|a(x_1,x_2;E|^2$ and then integrate over $x_1$ and $x_2$.
As a result, in the semiclassical approximation we obtain
$$\bar\o(E)=\lim_{j=0} \int_0^\infty \f{Dx Dp}{|Z_T|^2}
e^{-iS_0(p,x)} \d(E-H(p,x))\t$$\be\t\prod_t\d\le(\dot p+\f{\pa H}{\pa
x}-j\ri) \d\le(\dot x-\f{\pa H}{\pa p}\ri),\l{3.6}\ee where \be
S_0(p,x)=\lim_{t=0} (S_{T+t}(p,x)-S_{T-t}(p,x))\l{3.7}\ee can be
different from zero if the trajectory is a periodic one (note that in
the preceding section the time $T$ was fixed and therefore $S_0=0$
independently of the type of the trajectory). In (\r{3.6}) we have
also taken into account that $\pa_T S_T=-H$.

In order to calculate the remaining integrals in (\r{3.6}) we must
find all solutions of the equations \be\dot p+\f{\pa H}{\pa x}=j,~~
\dot x-\f{\pa H}{\pa p}=0\l{3.8}\ee in the vicinity of the point
$j=0$. First, for the potential chosen these equations have the
"trivial" solution \be x_1(t|j=0)=0,~~p_1(t|j=0)=0,\l{3.9}\ee which
in the semiclassical approximation corresponds to a particle at rest
at the bottom of the potential well. Expanding $x_1(t|j)$ and
$p_1(t|j)$ in powers of $j(t)$ [under the condition (\r{3.9})], we
obtain an expansion in powers of the nonlinearity of the potential
$V(x)$. It is not difficult to see that this expansion corresponds to
the standard perturbation theory (the proof of this statement in the
field theory will be given in our next paper) and describes the
Brownian motion of the particle under the influence of the perturbing
force $j(t)$. Below this contribution to $\bar\o(E)$ will be denoted
by $\bar\o_1(E)$.

Another solution of the equation (\r{3.8}) is a purely periodic
trajectory $(x_2,p_2)=(x_c,p_c)+O(j)$, where the orbit $(x_c,p_c)$
[not perturbed by the source $j(t)$] is an exact nontrivial solution
of the equations \be\dot p+\f{\pa H}{\pa x}=0,~~ \dot x-\f{\pa H}{\pa
p}=0\l{3.10}\ee of classical mechanics. This equations are
translationally invariant. Therefore \be x_c=x_c(t+t_0,\e),~~
p_c=p_c(t+t_0,\e),\l{3.11}\ee where $\e$ is the energy of the
particle on the trajectory $(x_c,p_c)$: \be H(x_c,p_c)=\e.\l{3.12}\ee
The integration in (\r{3.6}) is performed over all the trajectories,
which implies integration over $\e$ and $t_0$ as well (recall that
$T$ is the proper time). In other words, $$\bar\o_2(E)=\int
dt_0\d\e\d(E-\e)\int_0^\infty dT\exp\{-iS_0(p_c,x_c)\}\t$$\be\t \int
\f{Dx Dp}{|Z_T|^2} \prod_t\d\le(\dot p+\f{\pa H}{\pa x}\ri)
\d\le(\dot x-\f{\pa H}{\pa p}\ri).\l{3.13}\ee Here \be|Z_T|^2=\int Dx
Dp\prod_t\d(\dot x -p)\d(\dot p).\l{3.14}\ee

Since we are considering a "two-particle" problem with a potential
independent of time, we can always make a canonical transformation
$(x,p)\to(X,P)$ \be DxDp=DXDP,\l{3.15}\ee such that \be \dot
P=-\f{\pa H'}{\pa X}=0,~~\dot X=\f{\pa H'}{\pa P}=const,\l{3.16}\ee
where $H'$ is the transformed Hamiltonian. The new variables have the
meaning of the action-angle variables. Upon such a transformation,
taking into account the fact that $(x_c,p_c)$ ia a periodic
trajectory, we obtain \be\bar\o_2(E)=\int dt_0\d\e\d(E-\e)\int_0
^\infty dT\P_T\exp\{-iS_0(p_c,x_c)\},\l{3.17}\ee where $\P_T$ is a
phase factor which takes into account the periodicity of the
contribution \C{ae}.

Next, because $H$ has a constant sign, for periodic trajectory one
has \be S_)(p_c,x_c)=\int_T^Tdt p_c\dot x_c=\oint_{x_c}pdx\l{3.18}\ee
[which takes into account the uncertainty in taking the limit $t\to0$
in (\r{3.7})]. In the formula (\r{3.18}) \be
p=\pm[2(\e-V(x))]^{1/2}.\l{3.19}\ee Since $x_c$ is a periodic
function, one can write \be\int_0^\infty dT f_T(x_c)=\sum_{n=0}
^\infty\int_0^{T_1} dT f_{T+nT_1}\l{3.20}\ee  where $T_1$ is the
period: \be T_1(\e)=2\int_{z_-}^{z_+}\f{dz}{[2(\e-V(z)]^{1/2}},~~
V(z_\pm)=\e.\l{3.21}\ee Taking into account (\r{3.20}) and the fact
that $\oint pdx$ is independent of the integration path, we have \be
\oint_{z_c(T+nT_1)}pdx=\pm2n\int_{z_-}^{z_+}dz[2(\e-V(z)]^{1/2}\equiv
\pm nS_1(\e).\l{3.22}\ee As a result, from (\r{3.17}) we find that
\be\bar\o_2(E)=\O T_1(E)\sum_{n=0}^\infty(-1)^n\le(e^{-iS_i(E)n}+
e^{iS_i(E)n}\ri),\l{3.23}\ee where \be\O=\int dt_0\l{3.24}\ee is the
volume of the translational group of the Hamiltonian $H$. The
formulae (\r{3.23}) also takes into account that $\P_{nT_1}=(-1)^n$.

In (\r{3.23}) we have to complete the definition of the sum over $n$
by the standard prescription $E\to E-i0$. Then \be\bar\o_2(E)=\O
T_1(E)\le\{\f{1}{1+e^{-iS_1-0}}+\f{1}{1+e^{iS_1-0}}\ri\}=
\O\sum_n\d(E-E_n),\l{3.25}\ee since [see the formulae (\r{3.21})] $
\pa S_1(E)/\pa E=T_1(E)$; $E_n$ is defined by the usual
Bohr-Sommerfeld quantization rule \be S_1(E_n)=2\int_{z_-}^{z_+}
dz[2(E_n-V(z))]^{1/2}\equiv \pi(2n+1). \l{3.26}\ee

Summing the formulae obtained, from (\r{3.6}) we obtain \be\bar\o(E)=
\bar\o_1(e)+\O\sum_{n=0}^\infty\d(E-E_n)=\O\sum_{n=0}^\infty\d(E-E_n)(1
+ O(1/\O)),\l{3.27}\ee which demonstrate the dominance of the
contribution of the periodic trajectories. In other words, we have
shown that the standard periodic boundary conditions for the
Schrodinger equation for a particle in a potential well select the
only probable forms of motion.

\section{Selection rules}\0

Let us generalize the above result to systems with a large number of
degrees of freedom. If the trajectory fills densely the
$2N$-dimensional phase space volume, then by virtue of the invariance
of the equations of motion and of the measure $D^Nx D^Np$ under
canonical transformations, the quantity \be \int d^Nx_1d^Nx_2
W(x_2,T;x_1,0)=\O^{2N}\l{4.1}\ee coincides with the number of states
in the phase space. This follows from the fact that one can always
perform a canonical transformation [being in the frame of
semiclassical approximation] as a result of which the dependence on
the initial conditions disappears. Here it is important that the
integration is performed over all trajectories, which differ also in
their initial condition (see Sec.2). This demonstrates the dominance
of the ergodic fluxes in the phase space for quantum mechanical
problems.

Let us discuss the role of the conservation laws. If the system is
integrable, i.e. if there exist $N$ first integrals of motion, then
in the $@N$-dimensional phase space the system occupies a smaller
volume: when one makes the canonical transformation to the
action-angle variables, the trajectory wraps around the surface of an
$N$-dimensional hypertorus. Then, repeating the arguments above, we
find that the corresponding probability is \be \sim \O^N\l{4.2}\ee
The system with $2N=2$ degrees of freedom studied in Sec.3 is fully
integrable (in the semiclassical approximation), there is one
integral of motion (the energy), and therefore the probability is
$\sim\O^1$. In quantum mechanics the dominant motion is the motion in
the resonant tori, as the result of which the energy is quantized.
Finally, if the Hamilton equations have only the trivial solutions
$x(t|j=0)=0$, $p(t|j=0)=0$, i.e. if the motion is made possible only
as a result of quantum perturbations, then the probability is \be
\sim\O^0,\l{4.3}\ee since in the semiclassical approximation the
volume that the trajectory occupies in the phase space is zero.

Therefore, taking into account the definition of the
quantum-mechanical probability one can establish a one-valued
correspondence between the classical and quantum and, thus, introduce
the classical definition of measure into quantum theory. Here we note
an advantage of the approach based on functional integration.

Then in our approach the problem of quantization of an arbitrary
Lagrange theory is divided into two parts:

(a) one must know exactly all the regular solutions of the equation
of motion (the solutions must be regular since the definition of
probability implies the reversibility of motion ant therefore in the
variation of action the surface terms must be discarded; the solution
have to be exact since there is complete cancellation of the
contributions of the trajectories that are unphysical for the
non-homogeneous equation of motion);

(b) from the sum of the contributions of all such solutions one must
select those that lead to the motion (in the classical sense) in the
largest volume of the phase space, i.e. which are realized on the
largest measure.

\section{Conclusion}

Let us state the main results of the paper.

1. All the calculations must be performed in the space with the
Minkowski metric. This condition is important in the field theories
with a high group symmetries (such as the theories of the Yang-Mills
type) since for such theories one has yet not been able to perform
adequately the analytic continuation into the Euclidean region. (The
fact that a theory must satisfy certain conditions upon analytic
continuation in time is clear from \C{af}.) Apart from that, in the
pseudo-Euclidean metric one is able to take into account external
conditions with nontrivial time dependence without any restrictions.

2. The quantization can be performed without the transition to the
canonical formalism (see Sec.2), remaining in the Lagrange formalism
which is a more natural formalism for relativistic field theories.

3. In obtaining the contributions to the functional integrals onle
the exact solutions of the equation of motion must be taken into
account.

4. The contributions to functional integrals are found by variation
of the classical nonrenormalized action, which simplifies the
calculations considerably. This important feature of our approach is
discussed in forthcoming publications in the light of the phenomenon
of spontaneous symmetry breaking.

\vskip 0.4cm {\large \bf Acknowledgement}

To conclude, the author thanks E.M.Levin, L.N.Lipatov,
I.V.Paziashvili, I.G.Svimo\-nishvili, L.A.Slepchenko, and
Ya.A.Smorodinskii for useful advice and discussions.

{\footnotesize Translated by Gregory Toker}
\end{document}